\input phyzzx.tex

\twelvepoint


\def\bs{{\bar s}}
\def\ba{{\bar a}}
\def\bu{{\bar u}}
\def\bz{{\bar z}}
\def\ds{|\partial s|^2}
\def\dsb{|{\bar \partial}s|^2}
\def\det{{\rm det}}
\def\unit{{\rm I\!I}}


\REF\Witten {E. Witten, Nucl. Phys. {\bf B500} (1997) 3, hep-th/9703166}
\REF\HLW{P.S. Howe, N.D. Lambert and P.C. West, Phys. Lett. {\bf B} (1998), 
hep-th/9710034}
\REF\three{P.S. Howe, N.D. Lambert and P.C. West, Phys. Lett. {\bf B419} 
(1998) 79, hep-th/9710033}
\REF\LW{N.D. Lambert and P.C. West, Nucl. Phys. {\bf B524} (1998) 141, 
hep-th/9711040}
\REF\LWtwo{N.D. Lambert and P.C. West, Phys. Lett. {\bf B424} (1998) 281, 
hep-th/9801104}
\REF\SW{N. Seiberg and E. Witten, Nucl. Phys. {\bf B426} (1994) 19, 
hep-th/9407087}
\REF\BF{F. Ferrari and A. Bilal, Nucl. Phys. {\bf B469} (1996) 387, 
hep-th/9602082}
\REF\CM{C.G. Callan. Jr.  and J.M. Maldecena, Nucl. Phys. {\bf B513} (1998) 
198,  hep-th/9708147}
\REF\one{P.S. Howe, N.D. Lambert and P.C. West, Nucl. Phys. {\bf B515} (1998) 
203, hep-th/9709014}
\REF\Gibbons{G.W. Gibbons, Nucl. Phys. {\bf B514} (1998) 603, hep-th/9709027}
\REF\BGT{E. Bergshoeff, J. Gomis and P.K. Townsend, 
Phys. Lett. {\bf B421} (1998) 109, hep-th/9711043}
\REF\GP{J. Gutowski and G. Papadopoulos, Phys. Lett. {\bf B432} (1998) 97, 
hep-th/9802186}
\REF\FS{A. Fayyazuddin and M. Spalinski, Nucl. Phys. {\bf B508} (1997) 219, 
hep-th/9706087}
\REF\HY{M. Henningson and P. Yi, Phys. Rev. {\bf D57} (1998) 1291, 
hep-th/9707251}
\REF\Mikhailov{A. Mikhailov, {\it BPS States and Minimal Surfaces}, 
hep-th/9708068}
\REF\HSW{P.S. Howe and E. Sezgin, Phys. Lett. {\bf B394} (1997) 62,
hep-th/9611008; P.S. Howe, E. Sezgin and P.C. West, Phys. Lett. {\bf B399}
(1997) 49, hep-th/9702008; P.S. Howe, E. Sezgin and P.C. West, 
Phys. Lett. {\bf B400}
(1997) 255, hep-th/9702111}
\REF\GLW{J.P. Gauntlett, N.D. Lambert and P.C. West, {\it Supersymmetric
Fivebrane Solitons}, hep-th/9811024}
\REF\GLWold{J.P. Gauntlett, N.D. Lambert and P.C. West, hep-th/9803216, to
appear in Comm. Math. Phys.}
\REF\GPS{G. Gibbons, G. Papadopoulos and K. Stelle, 
Nucl. Phys. {\bf B508} (1997) 623, hep-th/9706207} 
\REF\Jose{J.M. Figueroa-O'Farrill, {\it Electromagnetic Duality for Children},
unpublished}
\REF\AH{M. Atiyah and N. Hitchin, {\it The Geometry and Dynamics of Magnetic
Monopoles}, Princeton University Press, Princeton, 1988}
\REF\KLVW{A. Klemm, W. Lerche, P. Mayr, C. Vafa and N. Warner, Nucl. Phys. 
{\bf B477} (1996) 746, hep-th/9604034}
\REF\SchW{J. Schulze and N. Warner, Nucl. Phys. {\bf B498} (1997) 101, 
hep-th/9702012}

\pubnum={KCL-TH-98-44  \cr hep-th/9811025}
\date{October 1998}

\titlepage

\title{\bf Monopole Dynamics from the  M-fivebrane}

\centerline{N.D. Lambert}

\centerline{and}

\centerline{P.C. West\foot{lambert, pwest@mth.kcl.ac.uk}}
\address{Department of Mathematics\break
         King's College, London\break
         England\break
         WC2R 2LS\break
         }

\abstract

We study the BPS states of the M-fivebrane which correspond to
monopoles of $N=2$ $SU(2)$ gauge theory.  Far away from the centres of the 
monopoles these states may be viewed as solitons in 
the Seiberg-Witten effective action. 
It is argued that these solutions are
smooth and some properties of their moduli space are discussed.



\endpage


\chapter{Introduction}

One of the most unexpected and detailed relations to come out the various
string theory dualities has been the connection between $p$-branes, viewed as
supergravity solitons, and Yang-Mills theories, obtained from the perturbative
D-brane description. 
In particular, this paper was motivated by the observation that a 
single M-fivebrane is capable of 
understanding some complex details of quantum non-Abelian gauge theory. 
More precisely  it was
observed in [\Witten] that the M-theory picture of $N$ D-fourbranes suspended
between two NS-fivebranes in type IIA string theory is just a single
M-fivebrane wrapped on a Riemann surface. It was then further argued 
that the Riemann surface in question is
none other than the Seiberg-Witten curve for the corresponding $N=2$
$SU(N)$ gauge theory  on the D-fourbranes. 

Let us consider this configuration from the M-fivebrane point of view.
We can denote this configuration by
listing the worldvolume directions of each M-fivebrane
$$
\matrix{M5:&1&2&3&4&5& & \cr
        M5:&1&2&3& & &6&7\cr}
\eqn\configone
$$
where we have suppressed the time dimension since this is  common to all
branes. To make contact with type IIA string theory and D-branes one
compactifies on $x^{7}$. In this case the first M-fivebrane becomes the two 
parallel NS-fivebranes and the second M-fivebrane  becomes  $N$
D-fourbranes.
From the point of view of the
first M-fivebrane the second M-fivebrane appears as a threebrane soliton
with worldvolume coordinates $(x^0,x^1,x^2,x^3)$. By solving the field 
equations for this configuration one sees that the threebrane can be viewed as 
simply the first  M-fivebrane wrapped on a Riemann surface [\three]. 
Moreover it was shown in 
[\HLW,\LW] that not just the elliptic curve but in fact 
the entire low energy Seiberg-Witten effective action can
obtained as the classical effective action for this threebrane soliton. 
Thus, a single M-fivebrane is capable of
predicting an infinite number of instanton coefficients in the four dimensional
non-Abelian gauge theory. Therefore one is naturally lead to the expectation 
that the M-fivebrane contains more information on 
non-Abelian fields than might naively be expected. 

To explore this possibility one is 
led to study the M-fivebrane  description of 
BPS states in Seiberg-Witten theory, or more precisely,  since we will obtain
results outside of the Seiberg-Witten effective description, low energy
$N=2$ super-Yang-Mills theory in the presence of BPS states. 
These states are of quite some interest and
have been studied from many points of view. Not least because of subtleties
in the predicted spectrum [\SW,\BF].
We will be particularly interested in
monopoles since these states are intimately connected to the non-Abelian gauge
structure and quantum dynamics. 

The configurations in question can be pictured as 
$$
\matrix{M5:&1&2&3&4&5& & \cr
        M5:&1&2&3& & &6&7\cr
        M2:& & & &4& &6& \cr
        M2:& & & & &5& &7\cr}
\eqn\configtwo
$$
Note that although there are two M-twobranes and two M-fivebranes there are
only three independent supersymmetry projectors, corresponding to $1/8$ of
spacetime supersymmetry or $1/4$ of the M-fivebrane worldvolume supersymmetry.
In other words after adding the first M-twobrane to the configuration 
\configone\ we find that the second M-twobrane can appear without breaking
any more supersymmetry. We include it to obtain the most general 
configuration. Indeed the appearance of two M-twobranes is crucial for our
analysis. 

One problem with self-dual strings obtained by intersecting an M-twobrane
with an M-fivebrane is that they have infinite tension, due to the infinite
length of the M-twobrane [\CM,\one,\Gibbons]. Clearly this is an unwanted
feature when trying to compare with the smooth BPS states in a 
Yang-Mills gauge theory. One way to avoid
the infinite energy is to place the M-twobrane suspended between two
parallel M-fivebranes. Then, despite the fact that the proper distance 
between the M-fivebranes diverges, the resulting
self-dual string has finite tension [\BGT]. Unfortunately, such a direct
approach is unavailable since  there is no known
description of two parallel M-fivebranes as this involves some kind of 
non-Abelian tensor multiplet. 

However, we wish to consider configurations with both 
self-dual strings and threebranes on the M-fivebrane worldvolume. In such a
situation  we expect to find finite energy solutions for the strings since 
they can be
stretched between the different branches of the same M-fivebrane. In other
words we wish to consider self-dual strings which are, in some sense,
wrapped around the Riemann surface which is embedded in spacetime. 
In this case the M-twobrane has a finite
worldvolume in spacetime and its boundary coincides with some cycle of the
Riemann surface. Furthermore the M-fivebrane worldvolume theory  
just contains an Abelian tensor multiplet. 

Another issue that we must
consider is whether or not the classical M-fivebrane equations provide an
accurate description of  the configuration
\configtwo, especially where the M-twobranes intersect the M-fivebranes. 
In order to ensure that the classical M-fivebrane's equations of motion 
are valid we must keep the curvatures small. We therefore need to look 
for smooth 
solutions in $x^1,x^2,x^3$ and $x^4,x^5$.
In addition we require that
the  space  derivatives $\partial_1,\partial_2,\partial_3$ are small. 
Indeed, following the
spirit of effective actions, we will only keep terms that are second order
in spacetime derivatives. The justification for this approximation, and 
the use of the classical M-fivebranes equations, therefore 
rests on the existence of well-behaved solutions to the effective equations 
of motion we will derive. 

It is natural 
to consider the resulting moduli space of these solutions
and compare it to that of monopoles in $N=2$ Yang-Mills theory. 
In fact we will argue below that there are finite energy solutions and 
that the two moduli spaces agree, thus providing an Abelian, M-theory 
description of monopole moduli space. This would mean that the low energy 
scattering of monopoles in Yang-Mills theory could be reproduced from
the M-fivebrane equations of motion describing self-dual strings scattering on
a Riemann surface. 
We note here that the   
moduli space of self-dual
strings  has been studied earlier in a different context [\GP]. 
There the infinite tension strings of [\HLW] 
(as obtained by the intersection of an 
M-twobrane with a single, flat  M-fivebrane) were considered and a 
hyper-K{\"a}hler with torsion metric was found.

The M-theory realisation of Seiberg-Witten theory and its BPS states  
have already been well studied in  [\FS,\HY,\Mikhailov]. However
these papers treated the M-fivebrane as infinitely heavy as compared to
the M-twobranes. It was therefore assumed that their geometry would be 
unaffected by the presence of the
self-dual strings. This assumption seems reasonable from the macroscopic
supergravity picture,  however, from the M-fivebrane
point of view the presence of self-dual strings dramatically alters the
geometry and even topology of the worldvolume [\CM,\one.\Gibbons]. 
Thus we may expect the
simple picture of an M-fivebrane wrapped around a Riemann surface, with 
M-twobranes attached, to break down near the self-dual strings. 
We therefore expect to see significant departures to the
Seiberg-Witten dynamics in the low energy effective action. Our approach then
offers a new method for studying these states in addition to the standard
field theory methods and the M-theory methods of  [\FS,\HY,\Mikhailov].

In the next section we discuss the Bogomoln'yi conditions and resulting
field equations for the M-fivebrane configuration \configtwo. In section
three we consider a limit where the equations of motion coincide with
what can be obtained from the Seiberg-Witten effective action. In the fourth
section we turn our attention to the full equations and argue for the 
existence of  smooth solutions and consider the moduli space. Finally 
we conclude with some comments in section five. 


\chapter{Self-dual Strings on a Riemann Surface}

\subsection{The Bogomoln'yi Equations}

In this paper we consider the worldvolume theory of an M-fivebrane with
coordinates $x^0,x^1,x^2,x^3,x^4,x^5$.  
We label the  six dimensional world indices by $m,n,p,...$ and four-dimensional
ones by $\mu,\nu,\rho,...=0,1,2,3$, of which we denote the three spatial
coordinates by $i,j,k,...=1,2,3$, or collectively as $\bf x$. 
Tangent indices are denoted by $a,b,c,...$. 
The M-fivebrane worldvolume theory possess a self-dual three-form 
$h_{mnp}$ and 
five scalar modes $X^6,...,X^{10}$ for bosonic fields and has eight fermion
degrees of freedom. The equations of motion are invariant under six
dimensional $(2,0)$ supersymmetry [\HSW] and an internal symmetry
$Spin(1,5)\times Spin(5)$ where $Spin(1,5)$ is the Lorentz group and
the $Spin(5)$ is an R-symmetry. 
The six-dimensional 
$\Gamma^a$-matrices are written in terms of matrices $\gamma^a$
$$
\Gamma^a = \left(\matrix{0&\gamma^a\cr \tilde\gamma^a&0}\right)\ , \quad
$$
with $\gamma^0=-\tilde\gamma^0=1$ and $\gamma^a=\tilde \gamma^a$, $a=1,...,5$
are five-dimensional Euclidean $\Gamma$-matrices.
In addition to these the worldvolume theory inherits a set of 
Euclidean $\Gamma$-matrices
from the five-dimensional space transverse to the M-fivebrane.
We denote these matrices by $\gamma_{a'}$, whose unique irreducible 
representation is a spinor transforming under the $Spin(5)$ R-symmetry.
It is important to note 
that the $\gamma^a$ and $\gamma_{a'}$ matrices act on different spinor
indices and so commute with each other.
The reader is  referred to [\HSW,\LW,\GLW] for more details of the notation. 

The configuration of an M-twobrane intersecting two M-fivebranes has also been
discussed in [\GLW] from the viewpoint of generalised calibrated geometries. 
Two intersecting  M-fivebranes in the 
$(x^0,x^1,x^2,x^3,x^4,x^5)$  
and $(x^0,x^1,x^2,x^3,X^6,X^7)$ planes reduce the supersymmetry to  
spinors such that [\one]
$$
\epsilon\gamma^{45}\gamma_{67} = -\epsilon\ .
$$
The configuration \configtwo\ also has an M-twobrane in
the $(x^0,x^5,X^7)$ plane with the corresponding projector [\three]
$$
\epsilon\gamma^{05}\gamma_7 = \eta\epsilon\ ,
$$
where $\eta = \pm 1$.
These two projectors actually imply that another M-twobrane can be 
introduced in the $(x^0,x^4,X^6)$ plane with the projector 
$\epsilon\gamma^{04}\gamma_{6} = -\eta\epsilon$, 
without breaking any additional 
supersymmetries. It is helpful to introduce complex notation 
$$
z = (x^4+ix^5)\Lambda^2\ ,
$$
where $\Lambda$ is a mass scale introduced for later convenience. 
The active M-fivebrane scalars are denoted by
$$
s = (X^6 +iX^7)/R\ ,
$$ 
where we treat $X^7$ as a compact coordinate with period $R$. Thus by taking
the limit of small $R$ we obtain a perturbative description in terms of
type IIA string theory as discussed in [\Witten].
For clarity
in this paper we will largely suppress the constants $\Lambda$ and $R$. 
In complex notation these projectors can simply be written as 
$$\eqalign{
\epsilon\gamma_{0z}&= \eta \epsilon \gamma_\bs\ ,\cr
\epsilon\gamma^z\gamma_s &=0\ .\cr}
\eqn\projectors
$$
In total this configuration preserves one quarter of the  M-fivebrane's
worldvolume supersymmetry, i.e. it preserves four supersymmetries,
the equivalent of $N=4$, $D=1$ supersymmetry.

Before proceeding with the equations of motion is it necessary to consider the
self-dual three-form $h_{abc}$ on the M-fivebrane worldvolume. 
This can be decomposed into a four-dimensional
vector $v_a$ and anti-symmetric tensor ${\cal F}_{ab}$ 
as follows (all indices are
in the tangent frame)
$$\eqalign{
h_{abz} &= \kappa{\cal F}_{ab}\ , \quad\quad 
h_{ab\bz}=\bar\kappa{\bar {\cal F}}_{ab}\ ,\cr
h_{az\bz} &= iv_a\ ,\cr}
\eqn\hansatz
$$
where self-duality implies that  $h_{abc} = 2\epsilon_{abcd}v^d$
and  ${\cal F}_{ab} = {i\over2}\epsilon_{abcd}{\cal F}^{cd}$. Here we
have introduced  $\kappa$ which 
is an arbitrary function and can be removed by a redefinition of 
${\cal F}_{\mu\nu}$. 
Later we will make
contact with the Seiberg-Witten effective action  where we write  
${\cal F}_{ab} = F_{ab}+{i\over2}\epsilon_{abcd}F^{cd}$. There we will 
need to choose $\kappa$ so that $F_{ab}$ satisfies 
the standard Bianchi identity. As discussed in [\HSW] 
$h_{abc}$ is not a closed three-form. Rather it is related to a closed
three-form $H_{abc}$ via
$$
H_{abc} = (m^{-1})_a^{\ d}h_{dbc}\ ,
\eqn\Hdef
$$
where $m_a^{\ b} = \delta_a^{\ b} - 2k_a^{\ b}$ and 
$k_{a}^{\ b}=h_{acd}h^{bcd}$. 
One can also derive the useful relation [\HSW]
$$
(m^{-1})_a^{\ b} = Q^{-1}(\delta_a^{\ b} + 2k_a^{\ b})\ ,
$$
where $Q = 1 - {2\over3}k_a^{\ b}k_b^{\ a}$. Finally we will use 
the worldvolume metric $g_{mn}$ which 
is just the standard induced metric
$$\eqalign{
g_{mn} &= \eta_{mn} + {1\over2}\partial_m s\partial_n\bs 
+ {1\over2}\partial_n s\partial_m\bs\ , \cr
&=e_m^{\ a}e_n^{\ b}\eta_{ab}\ .\cr}
$$

Since we are interested in the equations of motions at low energy we only
consider expressions up to second order in spacetime derivatives. We also
look for static solutions. 
For the convenience of the reader we list the components of the 
veilbein $e_{m}^{\ a}$,
to second order in spacetime derivatives, for the geometry used in this paper
$$\eqalign{
e_{\mu}^{\ a} & = \delta_{\mu}^{\ a} -{1\over2} 
\left(1\over \det e\right)^2  \left(
\bar\partial s\partial s\partial_{\mu}\bs\partial^a\bs 
+\partial \bs\bar\partial\bs\partial_{\mu}s\partial^a s\right)\ ,\cr
&\hskip1.1cm+{1\over4}\left({1+\ds+\dsb\over (\det e)^2}\right)
\left(\partial_{\mu}s\partial^{a}\bs+\partial_{\mu}\bs\partial^a s\right)\ ,\cr
e_{\mu}^{\ z} &= {(X^2-\ds)\bar\partial s\partial_{\mu}\bs
+ (X^2-\dsb)\bar\partial\bs\partial_{\mu}s\over X \det e}\ ,\cr
e_{\mu}^{\ \bz} &= {(X^2-\ds)\partial \bs\partial_{\mu}s
+ (X^2-\dsb)\partial s\partial_{\mu}\bs\over X \det e}\ ,\cr
e_z^{\ \bz} &= {\partial s\partial \bs\over X}\ ,\quad\quad
e_{\bz}^{\ z} =  {\bar\partial \bs\bar\partial s\over X}\ ,\cr
e_z^{\ z}& =
e_{\bz}^{\ \bz} = X\ ,\cr}
\eqn\veilbein
$$
where
$$\eqalign{
X^2 &= {1\over2}\left[(1+\ds+\dsb) + \det e\right]\ ,\cr
\det e  &= \sqrt{(1+\ds+\dsb)^2 - 4\ds\dsb}\ .\cr}
$$
For a more detailed discussion of the worldvolume fields and
equations of motion for the M-fivebrane we refer the reader to [\HSW].

The full non-linear supersymmetry variation for worldvolume fermions 
is derived in detail in [\GLW,\GLWold] 
$$\eqalign{
\hat \delta \Theta &=
{1\over 2}  \epsilon
\left\{ 
{\rm det}(e^{-1})\partial_m X^{c'}{(\gamma^m)}
{(\gamma_{c'})}\right.\cr
&\left. \hskip.75cm
- {1\over 3!} {\rm det}(e^{-1})\partial_{m_1} X^{c^\prime_1} \partial_{m_2}
X^{c^\prime_2}\partial_{m_3} X^{c^\prime_3}
(\gamma^{m_1m_2m_3})
{(\gamma_{c^\prime_1 c^\prime_2 c^\prime_3})}\right.\cr
&\left.\hskip.75cm
+{1\over 5!}{\rm det}(e^{-1}) \partial_{m_1} X^{c^\prime_1}\dots 
\partial_{m_5} X^{c^\prime_5} (\gamma^{m_1\ldots m_5})_{\alpha\beta}
{(\gamma_{c^\prime_1 \ldots c^\prime_5})}\right.\cr
&\left. \hskip.75cm
- h^{m_1m_2m_3}\partial_{m_2}X^{c_2'}\partial_{m_3}X^{c_3'}
(\gamma_{m_1})_{\alpha\beta}(\gamma_{c_2'c_3'})\right.\cr
&\left. \hskip.75cm
- {1\over3}h^{m_1m_2m_3}
(\gamma_{m_1m_2m_3})\delta_i^{\ j}
\right\}\ ,}
$$
where $\gamma_{mnp} = \gamma_{[m}\tilde\gamma_n\gamma_{p]}$ is anti-self-dual
and  we have adopted the convention that the $\gamma_m$ matrices always 
appear with tangent indices (i.e. $\gamma_m = \delta_m^a\gamma_a$). 
Specialising to the case with two scalars yields, in
complex notation,
$$
\hat\delta\Theta = 
\epsilon\left[{1\over2}{1\over \det e}\gamma^m\partial_m s\gamma_{s}
+{1\over2}{1\over \det e}\gamma^m\partial_m \bs\gamma_{\bs}
- {1\over2}h^{mnp}\gamma_m\partial_ns\partial_p\bs\gamma_{s\bs}
- {1\over3!}h^{mnp}\gamma_{mnp}\right]\ ,
\eqn\susy
$$
Note that the 
projectors \projectors\ imply that there are four 
independent terms appearing in \susy\  proportional to
$$
\epsilon\gamma_{0iz}\ ,\quad
\epsilon\gamma_{0z\bar z}\ , \quad 
\epsilon\gamma_{iz\bar z}\ , \quad
\epsilon\gamma_0\ , 
$$ 
and their complex conjugates.
Thus we may obtain the  Bogomoln'yi equations by setting the corresponding
coefficients  to zero. Using the decomposition \hansatz\ this yields
$$\eqalign{
\kappa {\cal F}_{0i} 
&= {1\over8}\eta\left({1+\ds-\dsb\over X^2 - \dsb}\right)
\left(
{X^2\partial_is + \partial\bs\partial s\partial_i\bs\over X \det e}
\right)\ ,\cr
v_0 &= +{i\over16}\eta\left({1+\ds-\dsb\over (X^2-\dsb)^2}\right)\left[
(1+\ds+\dsb){\bar\partial s\partial_is\partial^i\bs\over(\det e)^2}\right. \cr
&\left. 
\ \ \ +\dsb
{(\partial s\partial_i\bs\partial^i\bs-\bar\partial\bs\partial_is\partial^is)
\over (\det e)^2}\right]+ {i\over4}\eta{\bar\partial s\over X^2 - \dsb},\cr
v_i &= {1\over16}\eta\bar\partial s\left({1+\ds-\dsb\over (X^2-\dsb)^2}\right)
{\epsilon_{ijk}\partial^js\partial^k\bs\over \det e}\ ,\cr
{\bar \partial}s &= -\partial\bs\ ,}
\eqn\bogo
$$
respectively. 

Lastly we need to calculate the three-form $H$ which most naturally appears
in the equations of motion. To this end we calculate 
the matrix $m^{-1}=Q^{-1}(1+2k)$ as
$$
m^{-1} = Q^{-1}\left(\matrix{
\delta_{\mu}^{\ \nu}+2 k_{\mu}^{\ \nu}
&32i\bar\kappa v_0{\bar {\cal F}}_{\mu}^{\ 0}
&-32i\kappa v_0{{\cal F}}_{\mu}^{\ 0}
\cr
-16i\kappa v_0{{\cal F}}^{\nu 0}
&1-16v_0^2
&4\kappa^2{\cal F}^2
\cr
16i\bar\kappa v_0{\bar{\cal F}}^{\nu 0}
&4\bar\kappa^2{\bar{\cal F}}^2
&1-16v_0^2\cr
}\right)\ ,
$$
where
$$\eqalign{
k_{\mu}^{\ \nu}&=8v_0^2\delta_\mu^{\ \nu} + 16v_\mu v^\nu 
+4|\kappa|^2{\cal F}_{\mu\lambda}{\bar {\cal F}}^{\nu\lambda} 
+ 4|\kappa|^2{\bar {\cal F}}_{\mu\lambda}{\cal F}^{\nu\lambda}\ ,\cr
Q&= 1 - 256v_0^2(v_0^2 - 2|\kappa|^2{\cal F}_{0i}{\bar {\cal F}}^{0i})\ ,\cr}
$$
and then use the definition \Hdef.
Despite the complicated form of these Bogomoln'yi equations one finds
after a lengthy calculation that 
the three-form $H$ takes on a relatively simple form. In particular, in 
the world frame we find
$$\eqalign{
H_{iz\bz} &=0 \ ,\cr 
H_{0ij} &= 0\ ,\cr
H_{0iz} &=  {1\over 8}\eta\partial_i s\ ,\quad \quad
H_{0i\bz} = {1\over8}\eta\partial_i\bs\ ,\cr
H_{ijz} &=  {i\over 8}\eta\epsilon_{ijk}\partial^ks\ ,\quad\quad
H_{ij\bz} =  -{i\over8}\eta\epsilon_{ijk}\partial^k\bs \ ,\cr
H_{0z\bz} &= -{1\over4}\eta {\bar \partial}s\ , \cr
H_{ijk} &= -{i\over 8}\eta{\epsilon_{ijk}}\left({
4{\bar \partial}s+2{\bar \partial}s\partial_i s\partial^i\bs
+\partial s\partial_i\bs\partial^i\bs
-{\bar \partial}\bs\partial_is\partial^is \over 1+\ds-\dsb}
\right)\ .\cr}
\eqn\H
$$

Note that in contrast to the case with no self-dual strings, $s$ is no longer
a holomorphic function. Rather it only satisfies 
${\bar \partial}s = -\partial\bs$, which can 
be thought of as one of the Cauchy-Riemann equations (the other Cauchy-Riemann
equation is $\bar\partial s = \partial\bs$). This non-holomorphicity is not
unexpected in light of the
observations in [\HY,\Mikhailov] that the M-fivebranes wraps a 
Riemann surface with one complex structure while the M-twobrane wraps 
a Riemann surface with a different complex structure 
(note that the embedding space
${\bf R}^3\times S^1$ is hyper-K\"ahler and has three complex structures).
Here we see that the flux of $H$ through the surface $\Sigma$ 
measures the  non-holomorphicity   in  $s$, i.e. the self-dual strings
distort the holmorphic structure of the Riemann surface.

\subsection{The Equation of Motion}

Now we must turn our attention to solving the M-fivebrane's equation of
motion. Since the Bogomoln'yi equation relates the scalars to the three-form
we need only check the equation of motion for $H_{mnp}$. This in turn is
equivalent the to condition that $\partial_{[m}H_{npq]}=0$ [\HSW]. 
Most of the equations obtained from this condition 
are identically true on behalf of the Bogomoln'yi equations.
The only  non-trivial equation is $\partial_{[z}H_{ijk]}=0$ which yields
$$
\partial_i\partial^i s + R^2\Lambda^4\partial\left[{4R^{-2}{\bar \partial}s
+2{\bar \partial}s\partial_i s\partial^i\bs
+\partial s\partial_i\bs\partial^i\bs
-{\bar \partial}\bs\partial_is\partial^is \over 1+R^2\Lambda^4\ds-
R^2\Lambda^4\dsb}\right]=0\ ,
\eqn\seqn
$$
where we have reintroduced the constants $R$ and $\Lambda$ for 
future reference.
Alternatively 
$H_{mnp}$ is a closed three-form and admits a two form potential $b_{mn}$. 
A quick examination of {\H} leads to the following choice for $b_{mn}$.
$$\eqalign{
b_{0z} &= -{1\over8}\eta s\ ,\quad \quad b_{0\bz} = -{1\over 8}\eta\bs\ ,\cr
b_{ij} &= {1\over8}\eta\epsilon_{ijk} b^k\ ,\cr} 
$$
with all other components vanishing. Here $b^k$ is a vector field which 
must satisfy
$$
\partial b_k = i\partial_k s \ ,
\eqn\beqone
$$
and 
$$\partial^k b_k = -i\left[{4{\bar \partial}s
+2{\bar \partial}s\partial_i s\partial^i\bs
+\partial s\partial_i\bs\partial^i\bs
-{\bar \partial}\bs\partial_is\partial^is \over 1+\ds-\dsb}\right]\ .
\eqn\beqtwo
$$

In this paper we are interested in finding smooth solutions to \beqone\ and
\beqtwo\ (or equivalently \seqn) 
and discovering the data required to specify such a solution. 
From the eleven-dimensional picture  we wish to
consider self-dual strings which are wrapped around a cycle of the
manifold defined by the embedding $(z,\bz)\rightarrow (s,\bs)$. Let
us call this manifold $\Sigma$.
In particular we require that as either $|z|$ or $|{\bf x}|$ tends to
infinity, we move far away from the self-dual strings, i.e. 
$H\rightarrow 0$. Thus, on account of
\H, we therefore take
$$\eqalign{
\partial_is&\rightarrow 0\  {\rm as}\  {\bf x}\rightarrow\infty\ , \cr
\bar\partial s&\rightarrow 0\  {\rm as}\  |z|\rightarrow\infty\ .\cr}
$$
Let us now  expand $b^k$ and $s$ in power series for large $|{\bf x}|$ 
$$
b^k = x^k\sum_{n=0}^{\infty} {c_n \over |{\bf x}|^n}\ ,\quad
s = \sum_{n=0}^{\infty} {s_n \over |{\bf x}|^n}\ .
$$
As a consequence of \beqone\ and \beqtwo\ one 
can see from the ${\cal O}(|{\bf x}|^0)$ terms  
that $\bar\partial s_0 =0$. 
Thus as ${\bf x}\rightarrow\infty$
we are left with a holomorphic function $s_0(z)$. To make contact with
four-dimensional
$SU(2)$ gauge theory we impose  precisely the same
considerations as in [\Witten]. Namely, upon compactification on $X^7$ to 
type IIA string theory in ten dimensions, the M-fivebranes should
reduce to two parallel NS-fivebranes 
with two D-fourbranes suspended between them. The appropriate
curve $s_0$ is then [\Witten]
$$
e^{-s_0} = z^2-u \pm\sqrt{(z^2-u)^2 - 1}\ ,
\eqn\SWcurve
$$
where $u({\bf x})$ is the modulus of the curve and takes the value $u_0$ as
$|{\bf x}|\rightarrow \infty$.  Let us denote the resulting Riemann surface 
by $\Sigma_0$, which is of course just the Seiberg-Witten curve. We therefore
impose the boundary condition that $s \rightarrow s_0$ as 
$|{\bf x}|\rightarrow \infty$ where $s_0$ is given by \SWcurve.

The above boundary conditions impose smoothness of the solutions at
spatial infinity. 
Let us now look for smooth solutions to the equations of motion \beqone\
and \beqtwo\ in the interior. We will limit our discussion here to a single 
self-dual string depending
only on $z,\bz$ and $|{\bf x}|$. Since we want $H_{0iz}$ to be well defined
at the origin we will look for solutions  
with $\partial_i s\rightarrow 0$ as 
$|{\bf x}|\rightarrow 0$. In this case we find, 
for small $|{\bf x}|$,
$$
\partial^k b_k = -{4i\bar\partial s(0)\over 1+|\partial s(0)|^2
-|\bar\partial s(0)|^2}\ ,
$$
where $s(0)=s(z,\bz,0)$. If we now write $b^k = x^kb(|{\bf x}|)$
we find, near $|{\bf x}|=0$,
$$\eqalign{
b^k &= -{4\over 3}i x^k{\bar\partial s(0)\over 
1+|\partial s(0)|^2-|\bar\partial s(0)|^2}+{\cal O}(|{\bf x}|^3)\ ,\cr
s(z,\bz,|{\bf x}|) &= s(0) - {2\over3}|{\bf x}|^2\partial
\left[{\bar\partial s(0)\over
1+|\partial s(0)|^2-|\bar\partial s(0)|^2 }
\right]+{\cal O}(|{\bf x}|^4)\ .\cr} 
\eqn\sma
$$
Thus we find a smooth non-trivial solution, so long as $\bar\partial s\ne0$.
In the holomorphic case we shall see below that there is no finite
solution except $\partial_i s\equiv 0$. To show that there are indeed
solutions which are smooth everywhere requires that \sma\ 
can be extended smoothly 
to all $|{\bf x}|$ and furthermore that it 
satisfies the boundary conditions imposed above.

As an aside we note what happens when $\partial_is\equiv 0$, 
$\bar\partial s\ne 0$.
In this case the closure of $H$ leaves us with the condition
$$
{\bar\partial s\over 1+\ds-\dsb}= i\theta\ ,
\eqn\example
$$
where $\theta$ is any real constant. However, in contrast to the
solutions of interest in the rest of this paper, it follows from
\example\ that $\bar \partial s$ must
be everywhere non-vanishing. In addition one can check that although \example\ 
is a first order condition it automatically implies the second order
equation
$$\eqalign{
0&=g^{mn}\nabla_m\nabla_n s \cr
&=-{1\over 2}{1\over \det e}
\left\{\partial\left[{1-\ds+\dsb\over \det e}\bar\partial s\right]
+\bar\partial\left[{1+\ds-\dsb\over \det e}\partial s\right]\right\}  \ ,}
\eqn\eqexample
$$
which this is just the familiar 
minimal area equations for the surface defined by 
$s(z,\bz)$
$$
S = \int d^2 z\ \det e\ .
$$


\chapter{The Large Distance, Seiberg-Witten Limit}

In this section we will analyse the case ${\bar \partial}s=0$. 
Since in the large $|{\bf x}|$ limit, 
$s$ becomes a holomorphic function $s_0(z)$, the analysis in this section
applies to this asymptotic regime. More precisely we expect this to be a 
suitable approximation when $|{\bf x}| >> l$, where $l$ is the size of 
the cycles on the Riemann surface. In the interior 
$s$ will be non-holomorphic and the dynamics will differ significantly. 
However this Seiberg-Witten limit is sufficient to evaluate the charges
as seen at infinity.

For $\bar\partial s_0=0$ the equation of motion becomes simply
$$
\partial_i\partial^i s_0 + \partial\left[{
\partial s_0\partial_i\bs_0\partial^i\bs_0
-{\bar \partial}\bs_0\partial_is_0\partial^is_0 \over 1+|\partial s_0|^2}
\right]=0 \ .
$$
Furthermore from \SWcurve\ one can see that 
$\partial_is_0 = \partial_iu {ds_0/du} = \partial_i u 
\lambda_z$ where $\lambda = \lambda_z dz$ is the holomorphic one form on the 
Riemann surface $\Sigma_0$
described by $s_0$. We may follow the method of [\LW,\LWtwo] 
to reduce this equation
of motion to four dimensions. To be more precise we consider
$$
\int \left\{
\partial_i\partial^i s_0 + \partial\left[{
\partial s_0\partial_i\bs_0\partial^i\bs_0
-{\bar \partial}\bs_0\partial_is_0\partial^is_0 \over 1+|\partial s_0|^2}
\right]\right\}dz\wedge{\bar\lambda} = 0\ ,
$$ 
and express the resulting four dimensional equations in terms of the periods 
[\SW]
$$
a = \oint_A s_0 dz\ ,\quad\quad a_D = \oint_B s_0 dz\ ,
\eqn\adef
$$
where $A$ and $B$ are a basis of one cycles of the Riemann Surface.
The equation of motion can be expanded into the form
$$
\partial^i\partial_i u I + \partial^iu\partial_iu{dI\over du}
-\partial_iu\partial^iu J + \partial_i{\bar u}\partial^i{\bar u}K=0\ ,
$$
where we have introduced the integrals
$$\eqalign{
I &= \int \lambda\wedge {\bar \lambda} = 
(\tau-{\bar \tau}){da\over du}{d\ba\over d\bu}\ ,\cr
J &= \int  \partial\left({\lambda^2_z\bar\partial\bs_0 
\over 1+|\partial s_0|^2}\right)
dz\wedge\bar\lambda = 0 \ ,\cr 
K &= \int \partial\left({{\bar\lambda}^2_{\bz} \partial s_0
\over 1+|\partial s_0|^2}\right) 
dz\wedge\bar \lambda = 
-{d\bar \tau\over d\ba}\left({d\ba\over d\bu}\right)^3\ ,\cr}
$$
which where evaluated in [\LW,\LWtwo]. In this way we arrive at the
four dimensional equation of motion
$$
\partial_i\partial^ia (\tau-\bar\tau) + \partial^ia\partial_ia {d\tau\over da}
-\partial^i\ba\partial_i\ba{d\bar\tau\over d\ba} = 0 \ ,
$$
or more simply
$$
{\rm Im}\ \partial_i\partial^i  a_D = 0\ , \quad\quad
{\rm Im}\ \partial_i\partial^i  a = 0 \ .
\eqn\SWeqn
$$
These equations are nothing more than a special case of the equation of 
motion obtained from the Seiberg-Witten action
$$
S_{SW} = \int d^4x \ {\rm Im}\left( \tau \partial_{\mu}a\partial^{\mu}\ba 
+ 16\tau{\cal F}_{\mu\nu}{\cal F}^{\mu\nu}\right)\ ,
\eqn\SWaction
$$
and subsequently imposing the Bogomoln'yi condition 
$$
\partial_i a = {8}\eta{\cal F}_{0i}\ .
\eqn\bogoSW
$$
Comparing this with the first equation in \bogo\
$$
\partial_i s_0 = 8\eta \kappa {\rm det}(e){\cal F}_{0i} \ ,
$$  
for the case $\bar\partial s=0$ requires that 
$\kappa = {\rm det}(e^{-1}){ds_0\over da}$. Indeed this is precisely
the normalisation found in [\LW,\LWtwo] where it was  
needed to ensure that $F_{\mu\nu}$ is a curl and so can be identified 
with the field strength of a gauge field, namely the superpartner of the
scalar $a$. The
second equation in \SWeqn\ is then just the Bianchi identity for $F_{\mu\nu}$.

The Laplace equations \SWeqn\ have the general solution for point sources
given by
$$
{\rm Im}a = <{\rm Im}a> + \sum_n {Q_n\over |{\bf x} - {\bf y}_n|}\ , \quad\quad
{\rm Im}a_D = <{\rm Im}a_D> + \sum_n{P_n\over |{\bf x}-{\bf y}_n|}\ ,
\eqn\Imaad
$$
where $Q_n$ and $P_n$ are constants. 
Here ${\bf y}_n$ the centres of the solitons and
$<{\rm Im}a>$ and $<{\rm Im}a_D>$ are the imaginary parts of the vacuum 
expectation values of $a$  and $a_D$ respectively. 
In addition to \Imaad\ one needs the exact form for 
$a_D$ as a function of $a$ in order to
find ${\rm Re} (a)$ and ${\rm Re} (a_D)$. This can be obtained from \SWcurve\ 
and \adef\ leading to an expansion  [\SW]
$$
a_D = i{a\over \pi} + i{a\over \pi}{\rm ln}{a^2} 
+ i a\sum_{k=1}^\infty {c_k \over a^{4k}}\ ,
\eqn\expansion
$$
where the coefficients are $c_k$ are real.
From the Bogomoln'yi condition \bogoSW\ we find 
$$
E_i = {1\over 8}\eta\partial_i {\rm Re} (a)\ ,\quad  
B_i = -{1\over 8}\eta\partial_i{\rm Im} (a) 
\eqn\EB
$$
where $E_i = F_{0i}$ and $B_i = {1\over 2}\epsilon_{ijk}F^{jk}$. Thus magnetic
states correspond to imaginary $a$ and electric states to real $a$. 

For a magnetic state, where ${\rm Re} (a)$ is constant, the asymptotic
Magnetic field is precisely what one expects
$$
B^i = -{1\over8}\eta \partial^i {\rm Im} (a) = 
{1\over8} \eta \sum_n{Q_n(x^i - y_n^i)\over |{\bf x} - {\bf y}_n|^3} \ .
$$
From \expansion\ we may (in principle) determine $a_D$. Note that
the presence of the logarithm and the reality of the coefficients $c_k$ 
in \expansion\  imply that  
${\rm Im} (a_D)= -{\rm Im} (a)$, which is consistent with \SWeqn. One then
sees that the $Q_n$ are the magnetic charges of the monopoles.

The situation is rather different for electric states where ${\rm Im}(a)$ is
constant.
In this case the asymptotic form for ${\rm Re} (a)$ is determined by inverting
the function $a_D(a)$, which  is a very subtle system to solve. 
Noting that  $\partial_i{\rm Im}(a)=0$ and  $\tau = da_D/da$ we find 
$$
E^i = {1\over8}\eta\partial^i{\rm Re} (a) = -
{1\over8}\eta{1\over {\rm Im}\tau}
\sum_n{P_n(x^i - y_n^i)\over |{\bf x} - {\bf y}_n|^3}\ .
$$
Thus from the leading order behaviour at infinity 
we see that $-P_n/{\rm Im}\tau(<a>)$ can be 
identified with the electric charges of the solitons. However, the presence
of the non-trivial terms in $\tau$ alters the behaviour
of the electric fields in the interior, 
corresponding to the (anti-) screening of electric charge.

Finally we note that if we follow the Seiberg-Witten equations \SWeqn\
into the centre of  a soliton we find ${\rm Im} (a), {\rm Im} (a_D)
\rightarrow\infty$. 
Thus we are
pushed into a regime where only the perturbative terms (i.e. the first 
two terms in \expansion) in the effective
action are important. This presumably reflects the fact the the underlying
Yang-Mills theory is asymptotically free. 

We can evaluate the energy
of these configurations from \SWaction\ to be
$$\eqalign{
{\cal E}&= \int d^3x\ {\rm Im}\tau \left(\partial_i a\partial^i \ba 
+ 32E_iE^i+32B_iB^i\right) \cr
&= {3\over 2}\int d^3x \left(\partial_i{\rm Re} (a)\partial^i{\rm Im}(a_D)
- \partial_i{\rm Re} (a_D)\partial^i{\rm Im}(a) \right)\ ,\cr
& = {3\over2} \oint dS^i\left( {\rm Re} (a)\partial_i{\rm Im}(a_D)
-{\rm Re} (a_D)\partial_i{\rm Im}a\right)\ ,\cr}
\eqn\E
$$
where we have used the fact that the integrand is a 
total derivative to express $E$ as a surface integral. One can easily see that 
${\cal E}$ diverges due to the behaviour near the centres of the solitons. 
In particular, from \Imaad\ one
sees that the measures $dS^i\partial_i{\rm Im}(a)$ and 
$dS^i\partial_i{\rm Im}(a_D)$ 
remain finite near the centre of a soliton. However, because 
$|a|\rightarrow\infty$, only the first two terms in \expansion\ 
are important so that ${\rm Re}(a)$ and ${\rm Re}(a_D)$  
diverge, rendering ${\cal E}$ infinite.  In other words if we set
$\bar\partial s=0$ everywhere then we obtain solitons to the M-fivebrane
equations of motion 
which can be viewed as solutions of the Seiberg-Witten action but with
a divergent mass. Of course this divergence near the self-dual strings 
contradicts the low energy
approximation we have used and therefore does not accurately describe the
smooth BPS states of the M-fivebrane that we wish to study. 

One might think  that this divergence is specific to the form of the 
function $a_D(a)$
in the Seiberg-Witten solution. For example one might consider
the curves corresponding to field theories with a positive $\beta$-function. 
Let us suppose that instead of \SWcurve\ we had a curve  for which
$a_D(a)$ was finite as $a\rightarrow\infty$. There would therefore  
be no divergence in the energy at the soliton core. Such a theory would be 
unphysical 
however since 
$\tau = da_D/da$ and hence  $\tau (\infty)=0$. Thus effective
action would vanish at $a=\infty$ and furthermore the holomorphicity of 
$\tau$  would lead to a violation of  ${\rm Im}\tau \ge 0$. In fact 
this is impossible given the construction of $\tau$ as the period matrix of
a Riemann surface.
Alternatively, 
if one has a smooth solution then the energy $\cal E$ can be evaluated
by integrating over the sphere at infinity but this leads to an expression
which is not in general positive definite, contradicting the manifest
positive definite nature of $\cal E$.

Thus we find that by imposing the condition 
$\bar\partial s = 0$ 
we do not obtain smooth,
finite energy solutions. Indeed these solutions do not seem physical
from the point of view of the M-fivebrane/M-twobrane system either.
To see this consider a self-dual string which is wrapped around the $B$-cycle
of the Riemann surface (since $\bar\partial s=0$ there is a clear notion of
a Riemann surface). According to the solution \Imaad, as we approach the 
string at $|{\bf x}|=0$, $a$ diverges. Thus $u \propto a^2$ also diverges
which corresponds to the two D-fourbranes in the type IIA picture moving 
infinity far apart. On the Riemann surface this means that 
the length of the $B$-cycle grows without bound. In other words the
self-dual string is choosing to wrap around a cycle of infinite 
length. This helps to explain why the corresponding four-dimensional
solution has a divergent energy, even though we have argued in the introduction
that the self-dual string has a finite tension. 
Since the condition $\bar\partial s=0$ 
ignores most of the details of where the self-dual string lies on the Riemann
surface, it is perhaps not surprising that we obtain unphysical solutions.
In the next section we shall analysis the full
non-holomorphic equations, including the effects of self-dual strings. 
Intuitively one sees 
that there might be finite energy solutions since we would expect
the self-dual string to wrap around cycles with as small a length as possible.
Thus at $|{\bf x}|=0$ one might suppose that the cycle which the self-dual 
string wraps shrinks to zero size. In the Seiberg-Witten description these
are the strong coupling singularities where ${\rm Re} (a)$ and 
${\rm Re} (a_D)$ 
are finite, 
leading to a convergent form for ${\cal E}$ above. 
Of course these arguments are
highly speculative, nevertheless they are indications that finite energy
solutions do exist.


\chapter{Moduli Space}

We now wish to analyse the full equation of motion for self-dual strings
on a Riemann surface, i.e. without assuming that $s$ is holomorphic. As 
mentioned above we will use the boundary condition that at spatial infinity
$s$ just the Seiberg-Witten curve \SWcurve. The full equations of motion
then describe how $s(z,\bz)$ behaves as  a function of $\bf x$. 
Note that, even though
$s$ is no longer holomorphic, the the manifold $\Sigma$ is still topologically
a Riemann surface. This is because, by virtue of the boundary condition 
and the equation of motion, it is just a smooth (although 
non-holomorphic) deformation of the Seiberg-Witten curve. 
In other words we may still think of $\Sigma$ as a genus one 
Riemann surface, but one
that it is not embedded holomorphically in spacetime.

First we note from \seqn\ that, from the four-dimensional
point of view, the term
$$
\Lambda^4\partial\left[{4\bar \partial s\over 
1+R^2\Lambda^4\ds -R^2\Lambda^4\dsb}\right]
$$
acts as a source for $\partial_i\partial^is$. 
To help understand this term 
let us expand $s = s_0+R^2s_1$, where $\bar \partial s_0=0$, and consider  
the type IIA string theory description by letting $R\rightarrow0$,
keeping $R\Lambda^2$ fixed. 
The equation of motion for $s$ now becomes 
$$
\partial_i\partial^is_0 + R^2\Lambda^4\partial\left[{
\partial s_0\partial_i\bs_0\partial^i\bs_0 
- \bar\partial \bs_0 \partial_i s_0\partial^i s_0\over 
1+R^2\Lambda^4|\partial s_0|^2}\right] = 
-4R^2\Lambda^4\partial\left[\bar\partial s_1\over 1+
R^2\Lambda^4|\partial s_0|^2 \right]\ .
\eqn\snew
$$
We may proceed to obtain four-dimensional equations for
the BPS states as follows. We write \snew\  as
$$
E_z = \partial_i\partial^i s_0  + \partial(T - \bar T) 
+4R^2\Lambda^4\partial\left[\bar\partial s_1\over 
1+R^2\Lambda^4|\partial s_0|^2 \right]=0\ ,
$$
where 
$$
T = R^2\Lambda^4{\partial s_0 \partial_i\bs_0\partial^i \bs_0 \over 
1+R^2\Lambda^4|\partial s_0|^2}\ .
$$
Next we follow [\LW,\LWtwo] and consider
$$\eqalign{
0&=\int_A \left(E_z dz - \bar E_{\bz}d\bz\right)\cr 
&=\int_A \left( 
\partial_i\partial^is dz -  \partial_i\partial^i\bs d\bz + d(T-\bar T)
+ 4R^2\Lambda^4d\left[\bar\partial s_1\over 
1+R^2\Lambda^4|\partial s_0|^2\right]\right) \ .\cr}
$$
Here $A$ is the $A$-cycle of the Riemann surface defined by the curve 
$s_0(z)$. In particular  $s_0$ is given in \SWcurve. It is
therefore  possible to
chose the $A$-cycle to avoid any singular points of $T-\bar T$. Since $T$
is single valued
the contribution of $T$ in the four-dimensional equation of motion
vanishes and we arrive at the modified Seiberg-Witten equation
$$
\partial_i\partial^i{\rm Im}(a) = 2iR^2\Lambda^4\int_A d\left[
{\bar\partial s_1\over 1+R^2\Lambda^4|\partial s_0|^2}\right]\ .
\eqn\sourceA
$$
Similarly we may reduce over the $B$-cycle to obtain 
$$
\partial_i\partial^i {\rm Im}(a_D) = 2iR^2\Lambda^4\int_B d\left[
{\bar \partial s_1\over 1+R^2\Lambda^4|\partial s_0|^2}\right]\ .
\eqn\sourceB
$$
Clearly this method of dimensional reduction 
agrees with the one in section three when $s_1=0$.

In the above $\bar\partial s_1\propto H_{0z\bz}$ 
also appears as a total derivative. However, since we
do not  know the form of $\bar\partial s_1$ we can not say that it is
single-valued and hence that the integrals vanish. In fact  a self-dual
string wrapped around a Riemann surface acts as a domain wall in the
Riemann surface.
Thus in particular if the self-dual string wraps around the $B$-cycle then, 
as the $A$-cycle is traversed,  $H_{0z\bz}$ will increase by one unit of charge
when the self-dual string is crossed. Since the Riemann surface is compact, one
sees that $H_{0z\bz}$, and hence $\bar\partial s_1$, must be multi-valued. 
Therefore we expect that the $A$-cycle
integral will be non-zero. A similar situation occurs
for self-dual strings wrapped around the $A$-cycle. Thus we find there is
a source for ${\rm Im}(a)$  if the self-dual string wraps the $B$-cycle and
a source for ${\rm Im}(a_D)$  if the self-dual string wrap's the $A$-cycle.
In particular examining \EB\ shows that if the self-dual string wraps the 
$B$-cycle we obtain
magnetic sources and if it wraps the $A$-cycle we obtain electric sources.
This agrees with  previous studies [\FS,\HY,\Mikhailov] which 
identified the electric and magnetic
states as corresponding to self-dual strings wrapped around $A$ or $B$ cycles.

This is a detailed an analysis of the equations of motion 
that we have been able to obtain. In the above 
we have argued that there are smooth 
solutions to the equations of motion so let us now assume this to be the
case.  We can then, in principle, construct the low energy
equations of motion for the solitons by allowing their moduli 
to become time-dependent.  
The natural generalisation of \seqn\ to time-dependent configurations is
$$
\partial_\mu\partial^\mu s + \partial\left[{4{\bar \partial}s
+2{\bar \partial}s\partial_\mu s\partial^\mu\bs
+\partial s\partial_\mu\bs\partial^\mu\bs
-{\bar \partial}\bs\partial_\mu s\partial^\mu s \over 1+\ds-
\dsb}\right]=0\ .
\eqn\seqntwo
$$
In the case that $\bar\partial s=0$ one can check 
that this is indeed what the M-fivebrane equations
of motion yield by comparing with the equations of motion in [\LW]. 
However we have not checked this for the general case, although 
clearly it is the only possibility compatible with 
Lorentz invariance. 
After substituting in the general solution one would then integrate the 
equations of motion over the $x^i,z,\bar z$ coordinates.
The resulting  equations may also be viewed as arising from a 
one dimensional sigma model with the
the moduli space of solutions for a target space. 

Let us illustrate
this for the case $\bar\partial s=0$, even though for this case
we are not able to obtain smooth low energy behaviour.
Here one can follow precisely the
same steps for equation \seqntwo\ that we did for \seqn\ in the last section.
In this way we arrive at the four-dimensional equation
$$
\partial_\mu\partial^{\mu}a(\tau-\bar\tau)
+\partial^{\mu}a\partial_{\mu} a {d\tau\over da}
- \partial^{\mu}\bar a\partial_{\mu}\bar a{d\bar \tau\over d\bar a} = 0\ . 
$$
If we now substitute in the solution to the Bogomoln'yi equations \SWeqn\
with time-dependent moduli we obtain
$$
\ddot a(\tau-\bar\tau)
+{\dot a}^2{d\tau\over da}
- {\dot{\bar a}}^2{d\bar \tau\over d\bar a} = 0\ ,
$$
where a dot denotes a time derivative. These equations can now be viewed 
as arising from the effective action
$$
S = \int dt d^3x {\rm Im}\tau |{\dot a}|^2\ .
$$
The last step is to write $\dot a = \sum_\alpha 
{\partial a \over \partial y^\alpha}\dot y^\alpha$,
where $y^\alpha$ 
are the moduli, and integrate over space. If we had smooth solutions
this would then lead to an effective action
$$
S = \int dt g_{\alpha\beta}\dot y^\alpha \dot y^\beta\ .
$$
Here $g_{\alpha\beta}$ is an induced metric on the moduli space of
solutions. 
For the rest of this section we will try to consider
some properties of this metric, for the general case $\bar\partial s \ne 0$
where we expect smooth solutions to exist.

Since these solutions preserve one quarter of the sixteen worldvolume 
supersymmetries, this  sigma model must admit $N=4$, $D=1$ supersymmetry. 
However there are
two types of multiplet in one dimension with four supercharges. The first,
$N=4A$, 
is  essentially the dimensional reduction of two dimensional $(2,2)$
supersymmetry and requires that the moduli space metric is K\"ahler. The
second, $N=4B$, is related to the reduction of two-dimensional $(4,0)$ 
supersymmetry
and this requires that the moduli space metric is  Hyper-K\"ahler, or 
hyper-K\"ahler with torsion [\GPS]. Actually there is a subtlety here in that
the three complex structures need not be covariantly constant, hence they
need not be Hyper-K\"ahler in the strict sense of the word. 

First let us recall the situation for monopoles in $N=2$ super Yang-Mills 
gauge theory. It is well known that in a monopole background 
the only fermion zero modes are chiral in a certain Euclidean sense (see for
example [\Jose]).
Thus the supersymmetry is of the 
$N=4B$ type. This means that the moduli space
metric, which again must admit four supersymmetries, is hyper-K\"ahler
(possibly with torsion), rather
than just K\"ahler (although the complex structures need not be covariantly
constant). 
In fact
it is well known and can be proved directly that the monopole moduli space
metric is hyper-K\"ahler (in the strict sense of the word) [\AH].

To check the chirality of the preserved supersymmetries in our solution
we must construct
the four-dimensional $\Gamma$-matrices. Unfortunately, the reduction we 
considered in section two
to obtain six-dimensional $\Gamma$-matrices  is not very useful here.
Instead, 
let us  denote by $\Gamma_{(D)}$ the  $\Gamma$-matrices in $D$
dimensions. For $d$ even we can always consider the decomposition from 
$D$ to $d$ dimensions given by 
$$
\Gamma_{(D)} = \Bigg\lbrace\matrix{
\Gamma_{(d)}^a\otimes \unit & a=0,1,...,d-1\cr
\Gamma_{(d)}^{d+1}\otimes\Gamma_{(D-d)}^{a'} & a'=d,...,n-1\cr
}\ ,
$$
where $\Gamma_{(d)}^{d+1} =c \Gamma_{(d)}^{012...d-1}$ and $c$ is chosen
so that $(\Gamma_{(d)}^{d+1})^2=1$.
Thus for the case in hand we may set
$$
\Gamma_{(11)}^{0,...,5} = \Gamma_{(6)}^{0,...,5}\otimes\unit\ ,\quad\quad
\Gamma_{(11)}^{6,...,10}  = \Gamma_{(6)}^7\otimes\Gamma_{(5)}^{1,...,5}\ ,
$$
where $\Gamma_{(6)}^7= \Gamma_{(6)}^{012345}$. The Euclidean five-dimensional
$\Gamma$-matrices arise from the transverse space to the M-fivebrane and
can be further decomposed as
$$
\Gamma_{(5)}^{6,7} = \tau^{1,2}\otimes\unit\ ,\quad\quad
\Gamma_{(5)}^{8,9,10} = \tau^{3}\otimes\Sigma^{1,2,3}\ ,
$$
reflecting the presence of the second M-fivebrane. 
Next we can further reduce to four dimensions 
$$
\Gamma_{(6)}^{0,1,2,3} = \Gamma_{(4)}^{0,1,2,3}\otimes\unit ,\quad\quad
\Gamma_{(6)}^{4,5}  = \Gamma_{(4)}^5\otimes\sigma^{1,2}\ ,
$$
Here and above $\Sigma^i,\sigma^i$ and $\tau^i$, $i=1,2,3$ are 
three sets of Pauli matrices
and $\Gamma_{(4)}^5 = -i\Gamma_{(4)}^{0123}$. Under 
this decomposition the spinors $\epsilon$ now
carry four indices: $\epsilon^{\alpha,r,r',i}$, where $\alpha=1,2,3,4\ ,
r=1,2\ ,r'=1,2$ and $i = 1,2,3$.
The first index is just the four-dimensional $Spin(1,3)$ index. The last
index carries a representation of $SO(3)\cong SU(2)$ which can be
identified with the R-symmetry of the four-dimensional $N=2$ superalgebra
of the threebrane soliton.
The other two indices represent internal symmetries which are broken  by
the intersecting M-fivebranes. 

The projectors 
of the two M-fivebranes reduce the supersymmetries to  $N=2$ in four 
dimensions. In eleven dimensions the projections are
$\epsilon \Gamma_{(11)}^{012345} = \epsilon \Gamma_{(11)}^{012367} = \epsilon$,
which are expressed in four-dimensions as
$$
-\epsilon \Gamma_{4}^5\otimes\sigma^3\otimes\unit\otimes\unit =
-\epsilon \Gamma_{4}^5\otimes\unit\otimes\tau^3\otimes\unit= \epsilon\ ,
$$
respectively. Thus the four-dimensional $N=2$ supersymmetries have their
chirality correlated with internal $r,r'$ indices. 
Next we must construct the M-twobrane projector 
$\epsilon\Gamma_{(11)}^{057} = \eta\epsilon$, where again $\eta=\pm1$. In
four-dimensions this becomes
$$
-i\epsilon\Gamma_{(4)}^0
\otimes\sigma^1\otimes\tau^2\otimes\unit = \eta\epsilon\ .
$$
To see that this projects $\epsilon$ on to a set of chiral supersymmetries
we consider the Euclidean four-dimensional $\Gamma$-matrices defined by
$$
\bar \Gamma^{1,2,3} \equiv \Gamma_{(4)}^{1,2,3}
\otimes\unit\otimes\tau^2\otimes\unit\ ,\quad\quad
\bar \Gamma^{4} \equiv \Gamma_{(4)}^5\otimes\sigma^1\otimes\unit\otimes\unit\ .
$$
One can easily check that 
$\bar \Gamma^5 \equiv \bar\Gamma^{1234} = \Gamma_{(11)}^{057}$. Thus the
supersymmetries preserved by the solitons are chiral with respect to 
this Euclidean four-dimensional $\Gamma$-matrix algebra
$$
\epsilon \bar\Gamma^5 = \eta\epsilon\ .
$$
It then follows that the moduli space sigma model has
$N=4B$ supersymmetry in one dimension. Another way to see this is to note
that the preserved supersymmetries transform non-trivially under  
the  $SO(3)$ R-symmetry. This also forces the 
one-dimensional sigma model  
effective theory to have $N=4B$ supersymmetry.
By the same reasoning as above it follows that the moduli
space of solutions is again hyper-K\"ahler or hyper-K\"ahler with torsion.
However it seems reasonable to assume that, as in the Yang-Mills case,
there is no torsion and the complex structures are indeed 
covariantly constant. 
In addition, due to overall translational and rotational symmetries of
the configurations,  the soliton solutions constructed here have 
exactly the same symmetry properties as monopoles in the Yang-Mills theory, 
i.e., translational symmetry of the centre of mass and 
the action of the $SO(3)$ rotation group. We are therefore 
led to the conjecture that the two moduli space metrics agree. 

It follows from the $4B$ supersymmetry that their must be $4k$ bosonic
zero modes for a given soliton solution. Again this is precisely the same
as the number of  bosonic zero modes of a $k$-monopole in $SU(2)$ Yang-Mills
theory. As with monopoles it is clear that $3k$ of these zero modes come
from the locations of the centres of the solitons, i.e. the ${\bf y}_n$
in \Imaad. However, the other $k$ zero modes are less obvious. Their origin
is well-known though and they arise as non-trivial gauge transformations of
the vector field at infinity [\AH]. 

It is instructive to recall the role of these other zero modes in the standard 
treatment of monopoles in $N=2$ gauge theory. Namely dyonic states are obtained
by turning on their  conjugate momentum. Furthermore  these 
zero modes are periodic and hence  electric charge is discrete in the 
quantum theory. 
However, from the point of view of the M-fivebrane we have seen (see also
[\FS,\HY,\Mikhailov]) that  dyons correspond to wrapping the 
self-dual string around a combination of $A$ and $B$ cycles. This suggests
an interesting interpretation for the periodic zero modes in terms of the
geometry of the surface $\Sigma$.

In the situation considered here this
leads to a slight puzzle. Namely, if the one gets $k$ bosonic zero modes
from non-trivial gauge transformations at infinity, then what happened to
the fourth translational zero mode  of a self-dual string in six
dimensions? 
In fact it is not hard to see that the presence of the Riemann surface, i.e.
the two M-fivebranes, removes the fourth zero mode. More
precisely, a self-dual
string will only have a translational zero mode if there is an isometry in
a particular direction. However, the metric on the Riemann surface ensures that
there is a minimum length cycle, about which the self-dual string will wrap.
Moving the self-dual string off this cycle will then cost energy as the
string length is increased.

In the limit where $\bar\partial s=0$ we can  determine to some extent 
where the string must wrap. One can easily see from \SWcurve\ that there is a 
discrete
symmetry $s_0 \leftrightarrow -s_0$, corresponding to interchanging the two
NS-fivebranes in the type IIA picture. From the point of view of the Riemann
surface this corresponds to interchanging the two sheets which cover the plane.
Let us assume that there is a unique minimal length cycle, for each
homology class, which must 
therefore be 
invariant under this symmetry. Thus a minimum length 
curve must lie on both sheets of the $z$ plane and therefore,
since it is connected, it must pass through the branch cuts. For the 
$A$-cycle this means that the curve must run between the two branch points.
For the $B$-cycle one finds that, to respect the  
$s\leftrightarrow -s$ symmetry, the curve  must double-up unless it too 
passes through the branch points.
But the branch  
points $z=\pm\sqrt{u\pm1}$ are precisely the points where $s_0=0$.  
So we see that the self-dual string must
wrap a cycle running between the zeros of $s_0$. 
This interpretation of the zeros of the 
Seiberg-Witten
differential $s_0dz$ has arisen before within string theory studies of
the Seiberg-Witten solution [\KLVW,\SchW]. 


\chapter{Conclusion}

In this paper we have studied BPS states of the M-fivebrane which, under
type IIA/ M-theory duality, correspond
to monopole states in $N=2$ $SU(2)$ super-Yang-Mills theory. In particular
we discussed a differential equation for the solitons and the
relation of these solutions to Bogomoln'yi states in the Seiberg-Witten
effective theory. We saw that the M-fivebrane theory led to significant 
corrections to 
Seiberg-Witten dynamics and suggested the existence of smooth non-singular
solutions. Thus we argued that there is a smooth moduli space of solutions to
the M-fivebrane Bogomoln'yi equations. We also argued that the metric on this
space is hyper-K\"ahler and hence it is natural to relate it to the monopole
moduli space metric. Let us conclude now with some additional comments on
our work.

In general we have suppressed the dependence on the parameters 
$R$ and $\Lambda$.
Indeed for the case that $\bar\partial s=0$ the low energy dynamics are in fact
independent of both $R$ and $\Lambda$ [\Witten,\LW]. This is a crucial point 
which leads
to the expectation that the low energy dynamics are precisely the same as
for the perturbative Yang-Mills description [\Witten] (obtained from the 
type IIA string theory picture as $R\rightarrow 0$). In the 
$\bar\partial s\ne0$ case, however, one does see a non-trivial 
dependence on the parameter $R$.  However this also
leads to an extra parameter in the low energy theory which  may 
also enter into the moduli space metric.
We have argued that this moduli space has the same symmetries as
monopole moduli space. With the exception of the one and two monopole moduli
spaces, these symmetries do not uniquely specify the monopole metric [\AH]. 
Thus it is
possible that this extra parameter is associated to deformations of the
monopole moduli space which preserve  the symmetries.

A final point to consider is the stability of the BPS states. It was
shown in [\SW] that the spectrum of BPS states is non-trivial and indeed stable
BPS states can be made unstable as one varies the vacuum expectation value
$<a>$. In particular,  at weak coupling the theory contains dyons with
arbitrary integer electric charge and unit magnetic charge 
and the $W^{\pm}$ bosons. However 
at strong coupling only the monopole and dyon with unit electric charge 
are stable [\BF]. Note that in the case $\bar\partial s=0$ the BPS states are
given by self-dual strings wrapped around the Riemann surface 
[\FS,\HY,\Mikhailov]. 
Modular invariance of the Riemann surface
presumably leads to a complete $SL(2,{\bf Z})$ spectrum of monopole/dyon 
states.
However we have seen that in the full M-fivebrane description the Riemann
surface is no longer holomorphically embedded in spacetime and hence there 
is no $SL(2,{\bf Z})$ modular symmetry. Therefore the M-fivebrane does not 
immediately predict a full $SL(2,{\bf Z})$ spectrum of monopole/dyon states. 
It would be interesting to see if the correct BPS states can be
predicted by the M-fivebrane approach presented here. 

We would like to thank Jerome Gauntlett for helpful discussions.

\refout

\end